\documentclass[aps,pre,twocolumn,showpacs]{revtex4-1}
\usepackage[utf8]{inputenc}
\usepackage[english]{babel}
\usepackage{amsmath}
\usepackage{amssymb}
\usepackage{amsfonts}
\usepackage[dvips]{color}
\usepackage{graphicx}
\usepackage{subfigure}
\usepackage{epsfig}
\usepackage{pxfonts}
\begin{document}
\title{\bf Elastic Enhancement Factor:\\
from Mesoscopic Systems to Macroscopic Analogous Devices.}
\author{Valentin V. Sokolov}
\email{V.V.Sokolov@inp.nsk.su}
\affiliation{Budker Institute of Nuclear Physics of SB RAS -
Novosibirsk, Russia}
\affiliation{Novosibirsk Technical University - Novosibirsk,
Russia}
\author{Oleg V. Zhirov}
\affiliation{Budker Institute of Nuclear Physics of SB RAS -
Novosibirsk, Russia}
\affiliation{Novosibirsk State University - Novosibirsk, Russia}

\begin{abstract}
Excess of probabilities of elastic processes over inelastic ones is a characteristic feature of the chaotic resonance scattering predicted
by the random matrix theory (RMT). Quantitatively, this phenomenon is  characterized by the elastic enhancement factor 
%$F^{(\beta)}$ 
that is, essentially, a typical ratio of elastic and inelastic cross sections. Being measured experimentally, this quantity can provide important information on the character of dynamics of the complicated intermediate open system formed on the intermediate stage of various resonance scattering processes. We discuss properties of the enhancement factor in a wide scope from mesoscopoic systems as, for example, heavy nuclei to macroscopic electromagnetic analogous devices imitating two-dimensional quantum billiards. We demonstrate a substantial qualitative distinction between the elastic enhancement factor's peculiarities in these two cases. A complete analytical solution is found for the case of systems without time-reversal symmetry and only a few equivalent open scattering channels. 
\end{abstract}


\maketitle

\section{Introduction}
The elastic enhancement effect shows up in different kinds of resonance processes in nuclear and atomic physics, electron transport through quantum dots, or, transmission of electromagnetic waves through microwave cavities. Starting with Moldauer's pioneering papers \cite{Moldauer}, various aspects of this phenomenon have repeatedly attracted the attention of theorists as well as experimentalists \cite{collection_1, Lewenkopf, Pluhar, Verb, Fyod}.\\  

The subject-matter considered has gained, finally, a solid theoretical foundation in the random matrix approach to the problem of the chaotic resonance scattering that has been worked out in the fundamental paper \cite{Weiden}. This approach had made analytical calculations possible of the two-point scattering matrix correlation function that is a quantity of primary importance. In particular, the components relevant to the elastic enhancement problem are found to be:
\begin{equation}\label{C}
\begin{array}{c}
C^{\text{abab}}\left(\omega =\text{$\epsilon $t}_H\right) \equiv \langle S^{\text{ab}}\left(\frac{\omega}{2}\right) S^{\text{ab\,*}}       \left(-\frac{\omega}{2}\right)\rangle_{conn}
\\
\,\,\,\,\,\,\,\,\,\\
=\delta ^{\text{ab}}{T_a}^2\left(1-T_a\right)J^{(\beta)}_{\text{aa}}(\omega)
%\\
%\,\,\,\,\,\,\,\,\,\\
+\left(1+\delta _{\beta 1}\delta^{\text{ab}}\right)T_a T_b P^{(\beta)}_{\text{ab}}(\omega)\,,
\end{array}
\end{equation}
where the indexes $a$ and $b$ indicate the scattering channels with the transmission coefficients $T_a$ and $T_b$ and $\omega$ is a dimensionless energy displacement. At last, the superscript $\beta$ specifies the symmetry class $\beta=1$ in the case of systems with preserved time-reversal symmetry and $\beta=2$ if this symmetry is broken. The functions $J^{(\beta)}_{\text{aa}}(\omega)$ and $P^{(\beta)}_{\text{ab}}(\omega)$ are represented by the famous three- or twofold integrals that can be found in ref. \cite{Weiden} ($\beta=1$) and in the very instructive paper \cite{Fyod} (for both symmetry classes).\\
 
Finally, the elastic enhancement factor is defined as
\begin{equation}\label{F}
F=\frac{\sqrt{{\rm var\;} S^{aa}\times {\rm var\;} S^{bb} }}{{\rm var\;} S^{ab}}
\end{equation}
where the variances are equal to 
\begin{equation}\label{var_S}
{\rm var\;} S^{ab}=\langle|S^{ab}|^2\rangle-|\langle S^{ab}\rangle|^2=C^{\text{abab}}\left(\omega=0\right)\,.
\end{equation}
Basically, the enhancement factor $F$ could depend on the channel indexes through the transmission coefficients $T_a$. Such a possibility is excluded if all channels are statistically  equivalent: $T_a=T_b...=T$, which is what we suggest throughout this paper. Then the eqs. (\ref{C}, \ref{F}) reduce to
\begin{equation}\label{F_r}
F_M^{(\beta)}(T)=1+\delta _{\beta 1} +(1-T)\frac{J_M^{(\beta)}(T)}{P_M^{(\beta)}(T)}\,.
\end{equation}

\section{Verbaarschot's regime}

In the mesoscopic resonance collisions that involve intermediate %stage 
highly exited heavy nuclei or many-electron atoms with chaotic internal dynamics, a very large number $M\gg 1$ of very weak, $T\ll 1$, channels are typically open. In that case, the elastic enhancement factor depends on the only parameter $\eta=M T$ (Verbaarschot's regime, \cite{Verb}) and can be expressed \cite{Sokol_1} as 
\begin{equation}\label{F_var Q}
\begin{array}{c}
F^{(\beta)}(\eta)=1+\delta_{\beta 1}+\eta\,{\rm var\;} Q(\eta)\\
\,\,\,\,\,\,\,\,\\
=2+\delta_{\beta 1}-\eta\,\int_0^{\infty} d s\, e^{-\eta s}
\,B_2^{(\beta)}(s)\,,
\end{array}
\end{equation}\\
where 
\begin{equation}\label{var Q}
\begin{array}{c}
{\rm var\;} Q{(\eta)}=\frac{\langle Q^2\rangle}{\langle Q\rangle^2}-1
=\int_0^{\infty} d s\,e^{-\eta s}\left[1-B^{(\beta)}_2(s)\right]
\end{array}
\end{equation}\\
is the variance of the time delays\cite{Sokol_2}, whereas $B_2^{(\beta)}(s)$ stands for the Dyson's spectral binary form factor \cite{Mehta} belonging to the symmetry class $\beta$.\\

The dimensionless "openness" parameter $\eta=M T$ has \cite{Sokol_1} a clear physical meaning being the ratio $\eta=t_H/t_W$ of the two characteristic times: the Heisenberg time $t_H=\frac{2\pi\hbar}{d}$ and the dwell (or Weisskopf) time  $t_W=\hbar/\Gamma_W=\langle Q\rangle/T$. The first of them, $t_H$, is defined by the mean level spacing $d$ of the discrete energy spectrum of the Hermitian part $H$ of the total non-Hermitian effective Hamiltonian ${\cal H}$. It is the time that a spatially small wave packet of the incoming particle penetrating into the internal region needs to distinguish the discreteness of its spectrum. On the other hand, the dwell time is the time the incoming wave packet spends inside the internal region before escaping through a certain reaction channel. The enhancement factor is the more sensitive to the spectral fluctuations the longer, $t_W\gg t_H;\,\,\eta\ll 1$, the wave packet remains inside the interaction domain. Otherwise, $t_W\ll t_H;\,\,\eta\gg 1$, the enhancement factor carries no information on them at all.\\ 

The spectral form factor $B_2^{(\beta)}(s)$ looks very simple in the case of systems with broken $T$-symmetry \cite{Mehta}:
\begin{equation}\label{B_2^{(2)}}
B_2^{(2)}(s)=(1-s)\Theta(1-s).
\end{equation}    
No problem arises also with the subsequent $s$-integration
\begin{equation}\label{F^{(2)}}
F^{(2)}(\eta)=1+\frac{1-e^{-\eta}}{\eta}\,.
\end{equation} 
The task becomes appreciably more complicated in the presence of $T$-invariance. The spectral form factor reads now:  
\begin{equation}\label{B_2^{(1)}}
\begin{array}{c}
B_2^{(1)}(s)=\left[1-2s+s\ln\left(1+2s\right)\right]\Theta(1-s)\\
\,\,\,\,\,\,\,\,\\
+\left[s\ln\left(\frac{2s+1}{2s-1}\right)-1\right]\Theta(s-1)\,.
\end{array}
\end{equation} 
so the $s$-integration seems to be quite a problem. Surprisingly, it can be carried out in this case as well and results finally in the following remarkable relation:
\begin{equation}\label{F^{(1)}}
\begin{array}{c}
F^{(1)}(\eta)=F^{(2)}(\eta)+1\\
\,\,\,\,\,\,\,\,\\
-\left(\frac{1+\eta/2}{\eta}e^{-\eta}
-\frac{1-\eta/2}{\eta}\right)e^{\eta/2} Ei(-\eta/2)\,.
\end{array}
\end{equation}
To the best of our knowledge, this relation illustrated in Fig.1 has remained unknown up to now.\\ 
  
In both cases $\beta=1,\,2$, the enhancement factor monotonically decreases from $F^{(\beta)}(0)=2+\delta_{\beta 1}$ to $F^{(\beta)}(\infty)=1+\delta_{\beta 1}$ when the parameter $\eta$ increases. Being rather fast at the beginning, the descent of the factors $F^{(\beta)}(\eta)$ slows gradually down approaching their minimal values. At last, for both values of $\beta$ the slopes at the origin are identical:
\begin{equation}\label{F'}
\frac{d F^{(\beta)}(\eta)}{d \eta}\Big|_{\eta=0}= -\frac{1}{2}\,.
\end{equation}

\begin{figure}
\begin{center}
\includegraphics[width=7.cm,angle=0]{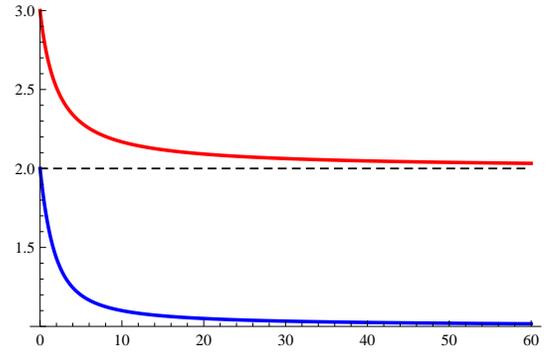}
\caption {$F^{(\beta)}$ versus $\eta$. Top curve, $\beta=1$, bottom curve, $\beta=2$.}
\label{fig:eta.}
\end{center}
\end{figure}

In spite of the obvious distinction of the expressions (\ref{F^{(2)}}) and (\ref{F^{(1)}}) their behavior is quite similar not only qualitatively but also quantitatively. The difference remains within a few percentage at most. (see Fig.2).
\begin{figure}
\begin{center}
\includegraphics[width=7.cm,angle=0]{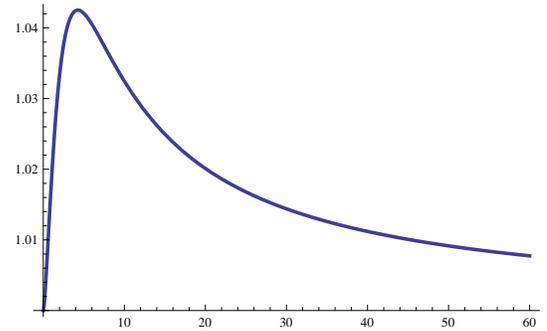}
\caption {The\, ratio\,\, $\frac{F^{(1)}}{F^{(2)}+1}$.}
\label{fig:ratio.}
\end{center}
\end{figure}

\section{Few channels}

Many aspects of the theory of quantum chaotic scattering can be
analyzed and checked experimentally with the aid of macroscopic analogous devices. This method took on a wide dissemination after pioneering experiments \cite{Stoek, Sridh} with open irregularly shaped two-dimensional (2D) electromagnetic resonators (see for example \cite{collection_2} and references therein). In particular, the elastic enhancement factor has thoroughly been measured for both symmetry classes as well as in the transient regime between them \cite{Dietz}. However, in contrast with the Verbaarschot's regime of very large number of very weak channels, the number $M$ of them is restricted in such experiments to only a few, as a rule even to two. The ruling parameter $\eta$ becomes irrelevant under such  conditions and the enhancement factor depends on $T$ and $M$ separately. 

\subsection{Broken time-reversal symmetry.}

According to Ref. \cite{Fyod}, the function $J_M^{(2)}(T)$ [see (\ref{F_r})] has the following twofold integral representation
($M\geq 2$):
\begin{equation}\label{J_M^{(2)}}
\begin{array}{c}
J_M^{(2)}(T)=\int_0^{\infty}\frac{d\lambda_1}{(1+\lambda_1 T)^{M+2}} \int_0^1{\left(1-\lambda T\right)}^{M-2}\,d\lambda\\
\,\,\,\,\,\,\,\,\\
=\frac{1}{T^2}\int_0^{\infty}\frac{dx}{(1+x)^{M+2}} \int_0^T\,(1-z)^{M-2}dz=\\
\,\,\,\,\,\,\,\,\\
=\frac{1}{T^2}\,\frac{1 -\,(1 - T)^{(M - 1)}}{M^2 - 1}\,.
\end{array}
\end{equation}
New variables $x=\lambda_1 T$ and $z=\lambda T$ have been introduced in the second line. In a similar way we then obtain 
\begin{equation}\label{P_M^{(2)}}
\begin{array}{c}
P_M^{(2)}(T)=\frac{1}{T^2}\int _0^{\infty }\frac{dx}{(1+x)^{M+2}}\\
\,\,\,\,\,\,\,\,\,\\
\times\int _0^T (1-z)^{M-2}\frac{T+x-z\,(1+ (2-T)\,x)}{x+z}dz\,.
\end{array}
\end{equation}
Unlike Eq. (\ref{J_M^{(2)}}), no general explicit formula exists in this case that would be valid for arbitrary number of channels $M$ and arbitrary value of the transmission coefficient $T$. Nonetheless, at any {\it fixed} value of $M$, the double integration (\ref{P_M^{(2)}}) can be fulfilled analytically. For instance,
\begin{equation}\label{P_{2,3}^{(2)}} 
P_2^{(2)}(T)=\frac{1}{6}+\frac{1}{3T}\,,\,\,\,\,\,
P_3^{(2)}(T)=\frac{1}{8}+\frac{1}{4T}-\frac{T}{24}\,,\,...
\end{equation}
Correspondingly, taking into account Eqs. (\ref{J_M^{(2)}} and \ref{F_r}) we arrive at 
\begin{equation}\label{F_{2,3}^{(2)}} 
F_2^{(2)}(T)=\frac{4-T}{2+T}\,,\,\,\,F_3^{(2)}(T)=\frac{2 \left(6-3 T+T^2\right)}{6+3 T-T^2}
\end{equation}
and so on. In such a manner, one can convince oneself that at {\it any given} number of channels $M$ the enhancement factor can be expressed, as a function of $T$, in the form of a ratio of two $(M-1)$-order polynomials. This statement is illustrated in Fig.3. It is clearly seen that the larger the number of channels the faster enhancement factor decays when $T$ increases. 

\begin{figure}
\begin{center}
\includegraphics[width=7.cm,angle=0]{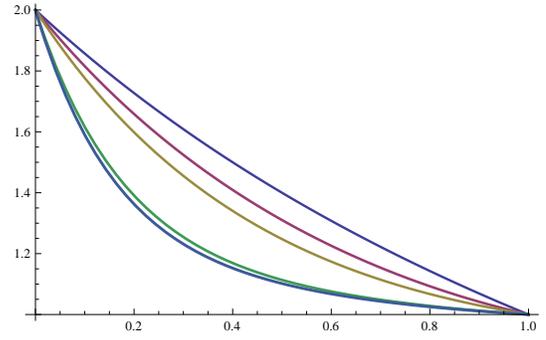}
\caption {$F_M^{(2)}$ versus $T$. From top to bottom $M=2,3,4,9,10.$}
\label{fig:T.}
\end{center}
\end{figure}

At last, it is worth mentioning that the connection established in Sec.II, between the enhancement factor and delay time variance, the latter being expressed now as ${\rm var\;}Q_M(T)=\frac{2}{T^2}\frac{1-(1-T)^{M+1}}{M^2-1}$ \cite{Somm},
does not exist anymore.\\

Further, we will restrict ourselves to the practically most interesting case of only two open channels. First, we would like to examine the significance of the assumption of equivalent channels. Let us suppose the opposite and define the following two new variables: 
$$T=\frac{1}{2}(T_1+T_2),\quad\Delta=\frac{1}{2}(T_1-T_2)$$
so\\
$T_1=T+\Delta,\,\,T_2=T-\Delta$ and $0<T<1,\,\,-\frac{1}{2}<\Delta<\frac{1}{2}$.\\

Though an explicit analytical expression still exists in this case, it turns out to be extremely lengthy. Therefore we skip the
formula and instead illustrate the result graphically (see Fig.4.). As long as $\Delta$ is noticeably smaller than $T$, the result is the same as in the case of equivalent channels and the latter approximation works well. Only when $\Delta$ is very close to $T$ the enhancement factor can become very large. The reason is quite simple: If either of the two channels is almost closed, almost everything is going through the only open one.\\

\begin{figure}
\begin{center}
\includegraphics[width=7.cm,angle=0]{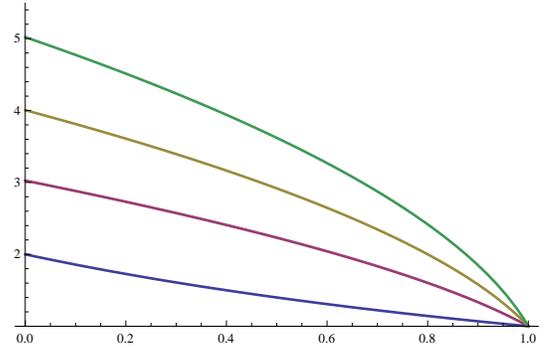}
\caption{$F(T+\Delta,T-\Delta)$ at different values of $\Delta$:
{from bottom to top ${\Delta=0,\, 0.994 T,\, 0.99985 T\,,0.9999985 T\,\,.}$
(Analytical solution.)}\label{fig:nequiv.}}
\end{center}
\end{figure}

%\vspace{0.6cm}
In the analogous experiments we discuss here the ohmic losses always play an important role and cannot be neglected. The simplest way of taking them into account consists in introducing the overall decaying factor $e^{-\gamma\tau}$ \cite{Fyod} where $\gamma$ plays the role of the absorption coefficient. In that case an explicit analytical expression $F_2^{(2)}(T,\gamma)=1+(1-T)R(T,\gamma)$ can still be found whereby the ratio $R$ is expressed as follows:
\begin{equation}
R(T,\gamma)\equiv\frac{J_2^{(2)}(T,\gamma)}{P_2^{(2)}(T,\gamma)}=
\frac{N(T,\gamma)}{D(T,\gamma)}\,,
\end{equation}
where the functions $D$ and $N$ are:

\vspace{0.3cm}
$N(T,\gamma)=T\,(2 T^2-T\gamma+\gamma^2)+\gamma^3\, e^{\gamma/T} Ei(-\gamma/T)\,,$

\vspace{0.3cm}
$D(T,\gamma)= T\left(T^3+\gamma^2-T\gamma(1+\gamma)+2T^2\left(2(\gamma-1)+
\frac{3\gamma}{e^{\gamma}-1}\right)\right)$\\
$+\gamma \left((1-T)\gamma^2-3\,T^2 \left(2-\gamma \coth(\gamma/T)\right)
\right) e^{\gamma/T} Ei(-\gamma/T)\,$.\\

The derived results are illustrated in the lower pallet of the Fig.5.
At the point $T=0$, the enhancement factor drops vertically down to the value 
\begin{equation}
F(T=0,\gamma)= 1+\frac{2}{\gamma \left(1+\text{Coth}\left[\frac{\gamma }{2}\right]\right)}
\end{equation}  
and, after that, approaches almost horizontally its minimal value $F(T=1,\gamma)=1$.\\

In the recent paper \cite{Sirko} results of the experimental investigation are reported of the properties of the elastic enhancement factor that have been carried out with the aid of analogue 2D electromagnetic resonators. The actual interval $5.2<\gamma<7.4$ of the absorption strength turned out to be appreciably larger that is shown in Fig.5. New calculations 
presented in \cite{Sokol_3} are in satisfactory agreement with
the experimental data. 

\subsection{Preserved time-reversal symmetry.}
Actually, the above-mentioned experiments \cite{Sirko} have been executed with a set up preserving time-reversal symmetry ($\beta=1$). No explicit analytical results can be derived in this case. Therefore we calculated the factor $F_{(M=2)}^{(\beta=1)}(T,\gamma)$ numerically in the same interval of the absorption coefficient to be able to compare the influence of the absorption in these two cases. The results that are presented in Fig.4. clearly demonstrate qualitative similarity between the two cases though, as it is seen, the T-invariant systems are somewhat more sensitive to the influence of the absorption.

%In the recent paper \cite{Sirko} results of the experimental investigation are reported of the properties of the elastic enhancement factor that have been carried out with the aid of analogue 2D electromagnetic resonators. The actual interval $5.2<\gamma<7.4$ of the absorption strength turned out to be appreciably larger that is shown in Fig.5. New calculations 
%presented in \cite{Sokolol_3} are in satisfactory agreement with
%the experimental data.

\begin{figure}
\begin{center}
\includegraphics[width=7.cm,angle=0]{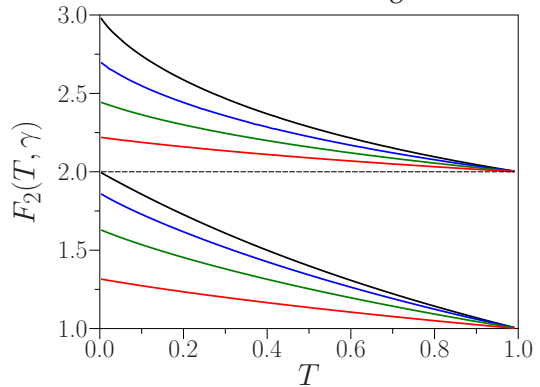}
\caption{$F^{(\beta)}_2(T,\gamma)$ for $\beta=1$ (up) and $\beta=2$
(down):\\
$\gamma=0,\, 0.03,\, 1,\, 3\,\,$(from top to bottom.)}\label{fig:absU&O.}
\end{center}
\end{figure}

%\vspace{0.3cm}
\section{Summary}
In this paper, we have concentrated our attention on the specific features of the elastic enhancement factor depending on the peculiarities of the chaotic open system with which one is dealing. On the whole, this factor depends on the number $M$ of scattering channels as well as the channel's transmission coefficients. However, when the number of channels is very large, what is typical of, for example, such processes as resonance nuclear reactions, the enhancement factor is controlled by the only parameter $\eta=M T$ that changes in very wide bounds (Verbaarschot's regime). Quite the opposite situation takes place in the analogous experiments with 2D irregularly shaped electromagnetic resonators that imitate quantum chaos. In these kinds of experiments the number of channels is very restricted. The enhancement factor depends on the number of channels and transmission coefficients separately in this case. We have juxtaposed in detail the two specified regimes. We have succeeded in finding a complete analytical solution valid for any fixed number $M$ of equivalent channels with arbitrary transmission coefficients $0<T<1$ in the case of systems without time-reversal symmetry. More than that, in the practically significant case of only two scattering channels, $M=2$, the influence of absorption is also described analytically. Finally, we have numerically demonstrated a close similarity between properties of the enhancement factors of systems with and without time-reversal symmetry.\\

{\bf Acknowledgments}
We are very grateful to L. Sirko for his interest in this work and significant critical remarks. V. Sokolov also appreciates useful discussions with J. Verbaarschot and F.M. Izrailev during the workshop on non-Hermitian Random Matrices``50 Yeas After Ginibre" (Israel 22-27 October 2014). This work was supported by the Ministry of Education and Science of the Russian Federation and by the Russian Foundation for Basic Research (Grant No. 14-02-00424). At last, we greatly appreciate the support from the RAS Joint Scientific Program ``Nonlinear Dynamics and Solitons".

\end{document}